\documentclass[conference]{IEEEtran}
\usepackage[left=0.67in,right=0.67in,top=0.64in,bottom=0.70in]{geometry}
\usepackage[T1]{fontenc}
\usepackage[latin9]{inputenc}
\usepackage{amsthm}
\usepackage{amsmath} 
\usepackage{bbm} 
\usepackage{graphicx} 
\usepackage{tabularx}
\usepackage{multirow, makecell}
\usepackage{color} 
\usepackage{cite}
\usepackage{algpseudocode}
\usepackage{algorithm}
\usepackage{amssymb}
\usepackage{subfigure}
\usepackage{esint}
\usepackage{epstopdf}
\usepackage{multirow}
\usepackage{makecell}
\usepackage{float}
\usepackage{lipsum}
\usepackage{balance}
\usepackage{upgreek}

\def\thickhline{\noalign{\hrule height.9pt}}

\newtheorem{definition}{Definition}
\newtheorem{remark}{Remark}

\newtheorem{notation}{Notation}

\theoremstyle{plain}
\theoremstyle{plain}
\newtheorem{theorem}{Theorem}
\newtheorem{lemma}{Lemma}

\newcommand{\comment}[1]{}

\IEEEoverridecommandlockouts

\allowdisplaybreaks

\begin{document}

\title{Probabilistic Design of Multi-Dimensional Spatially-Coupled Codes}

\author{
   \IEEEauthorblockN{Canberk \.{I}rima\u{g}z{\i}$^{1*}$, Ata Tanr{\i}kulu$^{2*}$, and Ahmed Hareedy$^{2}$}
   \IEEEauthorblockA{$^1$Institute of Applied Mathematics, Middle East Technical University, 06800 Ankara, Turkey \\
$^2$Department of Electrical and Electronics Engineering, Middle East Technical University, 06800 Ankara, Turkey\\
canberk.irimagzi@metu.edu.tr, ata.tanrikulu@metu.edu.tr, and ahareedy@metu.edu.tr}

\thanks{$^*$The first two authors have equal contribution to this work.} 
}
\maketitle

\begin{abstract}
Because of their excellent asymptotic and finite-length performance, spatially-coupled (SC) codes are a class of low-density parity-check codes that is gaining increasing attention. Multi-dimensional (MD) SC codes are constructed by connecting copies of an SC code via relocations in order to mitigate various sources of non-uniformity and improve performance in many data storage and data transmission systems. As the number of degrees of freedom in the MD-SC code design increases, appropriately exploiting them becomes more difficult because of the complexity growth of the design process. In this paper, we propose a probabilistic framework for the MD-SC code design, which is based on the gradient-descent (GD) algorithm, to design better MD codes and address this challenge. In particular, we express the expected number of short cycles, which we seek to minimize, in the graph representation of the code in terms of entries of a probability-distribution matrix that characterizes the MD-SC code design. We then find a locally-optimal probability distribution, which serves as the starting point of a finite-length algorithmic optimizer that produces the final MD-SC code. We offer the theoretical analysis as well as the algorithms, and we present experimental results demonstrating that our MD codes, conveniently called GD-MD codes, have notably lower short cycle numbers compared with the available state-of-the-art. Moreover, our algorithms converge on solutions in few iterations, which confirms the complexity reduction as a result of limiting the search space via the locally-optimal GD-MD distributions.
\end{abstract}


\section{Introduction}

Low-density parity-check (LDPC) codes \cite{gal_th} have been among the most widely-used error-correction techniques in several different applications. The performance of an LDPC code is negatively affected by detrimental configurations in the graph representation of the code, namely absorbing sets \cite{absorbing} formed by a combination of short cycles \cite{fossorier}. Spatially-coupled (SC) codes are a class of LDPC codes that offer excellent asymptotic performance \cite{kudekar, kudekar2}, superior finite-length performance \cite{costello, battaglioni, GRADE}, and low-latency windowed decoding that is suitable for streaming applications \cite{siegel}. Here, we focus on time-invariant SC codes \cite{battaglioni, amir}. SC codes are designed by partitioning an underlying block code into $m$ components, then coupling these components multiple times. The number of components $m$ is called the memory of the SC code.

The finite-length performance advantage of SC codes results from the additional degrees of freedom they offer the code designer via partitioning. Those additional degrees of freedom grow with the memory of the SC code. There are various results on high performance SC code designs with low memories in the literature \cite{battaglioni, rosnes, battaglioni-globe}. Some of these designs are optimal with respect to the number of detrimental objects \cite{oocpo}; others demonstrated notable performance gains in data storage and transmission systems \cite{hareedy-flash, channel_aware}. Optimal solutions are notorious to achieve for SC codes with high memories because of the remarkable complexity growth. Therefore, a recent idea introduced probabilistic design of SC codes with high memories, which is based on the gradient-descent (GD or GRADE) algorithm, to find locally-optimal solutions with respect to the number of detrimental objects \cite{GRADE}.

Multi-dimensional spatially-coupled (MD-SC) codes are a new class of LDPC codes that has excellent performance in all regions \cite{ohashi, truhachev-mitchell-lentmaier, liu}. These codes can mitigate various sources of (multi-dimensional) non-uniformity \cite{dahandeh, cai} in many applications. There are various effective MD-SC code designs introduced in recent literature \cite{schmalen, truhachev}. One idea is to use informed MD relocations to connect $M$ copies of a high performance SC code to construct a circulant-based MD-SC code \cite{homa-hareedy, homa-lev, rohith2}, which is the idea we focus on in this work. MD-SC codes offer additional degrees of freedom to the code designer. However, exploiting all of these degrees of freedom remains an open problem because of complexity constraints.

In this paper, we present the first probabilistic framework for the design of high performance MD-SC codes. Our framework enables MD-SC code designs with high memory $m$ and high number of SC copies $M$, implying an abundance of degrees of freedom. We use MD relocations to remove error-prone objects, and we focus here on short cycles, to improve performance. Our idea is to build a probability-distribution matrix, the rows of which correspond to the component matrices of the SC code, while the columns of which represent the auxiliary matrices of the MD code, which are used to connect SC copies. Each entry in this matrix represents the probability that a non-zero element will be associated with the corresponding component matrix at the corresponding auxiliary matrix in the MD-SC construction. Via our MD-GRADE algorithm, we find a locally-optimal distribution matrix, minimizing the expected number of short cycles. This locally-optimal distribution is the input to a finite-length algorithmic optimizer (FL-AO), which specifies the final MD-SC construction.

We develop the objective functions representing the expected number of short cycles in terms of the aforementioned probabilities. We obtain the required gradients of the objective functions and derive the form of the solution. We introduce the MD-GRADE and the FL-AO algorithms. Our FL-AO converges on excellent finite-length MD-SC designs in few iterations because of the guidance offered by the distribution obtained from the proposed GD analysis and algorithm. In particular, the GD-MD distribution results in a significant reduction in the search space the FL-AO operates on, which remarkably reduces the framework complexity and latency. Low relocation percentages are adopted because of decoding-latency restrictions \cite{siegel, homa-lev}. Furthermore, our experimental results demonstrate that the new codes, which we call GD-MD codes, have significantly lower populations of short cycles compared with the available state-of-the-art.

The rest of the paper is organized as follows. In Section~\ref{sec: prelim}, we present the necessary preliminaries. In Section~\ref{sec: framework}, we introduce the theory of our probabilistic framework for MD-SC code design. In Section~\ref{sec: algorithms}, we introduce the MD-GRADE and FL-AO algorithms. In Section~\ref{sec: experiment}, we present experimental results and comparisons against literature. In Section~\ref{sec: concl}, we conclude the paper.

\section{Preliminaries}\label{sec: prelim}

Let $\mathbf{H}_{\textup{SC}}$ in (\ref{sc matrix}) be the parity-check matrix of a circulant-based (CB) SC code with parameters $(\gamma, \kappa, z, L, m)$, where $\gamma \geq 3$ and $\kappa > \gamma$ are the column and row weights of the underlying block code, respectively. The SC coupling length and memory are $L$ and $m$, respectively.

\vspace{-0.7em}\begin{gather} \label{sc matrix}
\mathbf{H}_{\textup{SC}} \triangleq
\begin{bmatrix}
\mathbf{H}_0 & \mathbf{0}  &  & & & \mathbf{0} \vspace{-0.5em}\\
\mathbf{H}_1 & \mathbf{H}_0 & &  &  & \vdots \vspace{-0.5em}\\
\vdots & \mathbf{H}_1 & \ddots &  &  & \vdots \vspace{-0.5em}\\
\mathbf{H}_m & \vdots & \ddots & \ddots & & \mathbf{0} \vspace{-0.5em}\\
\mathbf{0} & \mathbf{H}_m & \ddots & \ddots & \ddots  & \mathbf{H}_0 \vspace{-0.5em}\\
\vdots & \mathbf{0} & & \ddots & \ddots & \mathbf{H}_1 \vspace{-0.5em}\\
\vdots & \vdots &  & &  \ddots & \vdots \vspace{-0.0em}\\
\mathbf{0} & \mathbf{0}& &  &  & \mathbf{H}_m
\end{bmatrix}.
\end{gather}
$\mathbf{H}_{\textup{SC}}$ is obtained from some binary matrix $\mathbf{H}^{\textup{g}}_{\textup{SC}}$ by replacing each nonzero (zero) entry in it by a circulant (zero) $z \times z$ matrix, $z \in \mathbb{N}$. Throughout the paper, the notation ``$\textup{g}$'' in the superscript of any matrix refers to its protograph matrix, where each circulant matrix is replaced by $1$ and each zero matrix is replaced by $0$. $\mathbf{H}^{\textup{g}}_{\textup{SC}}$ is called the protograph matrix of the SC code, and it has the convolutional structure composed of $L$ replicas $\mathbf{R}_{i}^{\textup{g}}$ of size $(m+L)\gamma \times \kappa$, where 
$$\mathbf{R}_1^{\textup{g}} = [(\mathbf{H}_0^{\textup{g}})^{\textup{T}} \textup{ } (\mathbf{H}_1^{\textup{g}})^{\textup{T}} \textup{ } \dots \textup{ } (\mathbf{H}_m^{\textup{g}})^{\textup{T}} \textup{ } (\mathbf{0}_{\gamma  \times \kappa })^{\textup{T}} \dots \textup{ } (\mathbf{0}_{\gamma  \times \kappa })^{\textup{T}}]^{\textup{T}}$$ and $\mathbf{H}_y^{\textup{g}}$'s are all of size $\gamma  \times \kappa$, $\gamma, \kappa \in \mathbb{N}$, and are pairwise disjoint. The sum $\mathbf{{H}^{\textup{g}}} = \sum_{y=0}^m \mathbf{H}_y^{\textup{g}}$ is called the base matrix, which is the protograph matrix of the underlying block matrix $\mathbf{H} = \sum_{y=0}^m \mathbf{H}_y$. In this paper, $\mathbf{{H}^{\textup{g}}}$ is taken to be all-one matrix, and the SC codes will have quasi-cyclic (QC) structure. 
\par
The $\gamma \times \kappa$ matrix $\mathbf{K}$ whose entry at $(i,j)$ is $a \in \{0,1,\dots,m\}$ when $\mathbf{H}_a^{\textup{g}}=1$ at that entry is called the partitioning matrix. The $\gamma \times \kappa$ matrix $\mathbf{L}$ with $\mathbf{L}(i,j)=f_{i,j} \in \{0,1,\dots,z-1\}$, where $\sigma^{f_{i,j}}$ is the circulant in $\mathbf{H}$ lifted from the entry $\mathbf{{H}^{\textup{g}}}(i,j)$, is called the lifting matrix. Here, $\sigma$ is the $z \times z$ identity matrix cyclically shifted one unit to the left. Observe that $m$ is the number of component matrices $\mathbf{H}_y^{\textup{g}}$ (or $\mathbf{H}_y$) and $L$ is the number of replicas in $\mathbf{H}^{\textup{g}}_{\textup{SC}}$ (or $\mathbf{H}_{\textup{SC}}$). 
\par
We now define the MD-SC code. The parity-check matrix $\mathbf{H}_{\textup{MD}}$ of the MD-SC code has the following form:
\vspace{-0.1em}\begin{gather} \label{md matrix}
\mathbf{H}_{\textup{MD}} \triangleq
\begin{bmatrix}
\mathbf{H}'_{\textup{SC}} & \mathbf{X}_{M-1} & \mathbf{X}_{M-2} & \dots & \mathbf{X}_{2} & \mathbf{X}_{1} \vspace{-0.0em}\\
\mathbf{X}_{1} & \mathbf{H}'_{\textup{SC}} & \mathbf{X}_{M-1} & \dots & \mathbf{X}_{3} & \mathbf{X}_{2} \vspace{-0.0em}\\
\mathbf{X}_{2} & \mathbf{X}_{1} & \mathbf{H}'_{\textup{SC}} & \dots & \mathbf{X}_{4} & \mathbf{X}_{3} \vspace{-0.0em}\\
\vdots & \vdots & \vdots & \ddots & \vdots & \vdots \vspace{-0.0em}\\
\mathbf{X}_{M-2} & \mathbf{X}_{M-3} & \mathbf{X}_{M-4} & \dots & \mathbf{H}'_{\textup{SC}} & \mathbf{X}_{M-1} \vspace{-0.0em}\\
\mathbf{X}_{M-1} & \mathbf{X}_{M-2} & \mathbf{X}_{M-3} & \dots & \mathbf{X}_{1} & \mathbf{H}'_{\textup{SC}}
\end{bmatrix},
\end{gather}
where 
\vspace{-0.5em}\begin{equation} \label{hmd constraint}\mathbf{H}_{\textup{SC}} = \mathbf{H}'_{\textup{SC}} + \sum_{\ell=1}^{M-1} \mathbf{X}_\ell.
\end{equation}  Here, $\mathbf{X}_\ell$'s are called the auxiliary matrices. The MD-SC code is obtained from $M$ copies of $\mathbf{H}_{\textup{SC}}$ on the diagonal by relocating some of its circulants from each replica of every copy of $\mathbf{H}_{\textup{SC}}$ to the corresponding locations in $\mathbf{X}_{\ell}$ copies (simply by shifting the circulants $\ell L\kappa z$ units cyclically to the left), and then by coupling them in a sliding manner as in (\ref{md matrix}). For convention, $\mathbf{X}_0 \triangleq \mathbf{H}_{\textup{SC}}'$.

\begin{definition} The $M (m+L)\gamma  \times M L \kappa $ matrix $\mathbf{H}^{\textup{g}}_{\textup{MD}}$ obtained by replacing $\mathbf{H}'_{\textup{SC}}$ with $\mathbf{H}'^{\textup{g}}_{\textup{SC}}$ and $ \mathbf{X}_\ell$'s with $ \mathbf{X}^{\textup{g}}_\ell$'s in (\ref{md matrix}) is called the MD protograph of $\mathbf{H}_{\textup{MD}}$. Here, $\mathbf{X}^{\textup{g}}_{\ell}$'s are uniquely determined by the following properties:
\begin{enumerate} 
\item The equation $\mathbf{H}^{\textup{g}}_{\textup{SC}} = \mathbf{H}'^{\textup{g}}_{\textup{SC}} + \sum_{\ell=1}^{M-1} \mathbf{X}^{\textup{g}}_\ell$ holds.
\item $\mathbf{X}_{\ell}$'s are obtained by replacing each nonzero (zero) entry in $\mathbf{X}^{\textup{g}}_{\ell}$ with the $z \times z$ circulant $\sigma^{f_{i,j}}$ (the zero matrix $\mathbf{0}_{z \times z}$), $z \in \mathbb{N}$, that has the appropriate power $f_{i,j}$ from the lifting matrix $\mathbf{L}$. 
\end{enumerate}
\end{definition} 

Relocations are represented by an MD mapping as follows: 
$$F: \{\mathcal{C}_{i,j} \,|\, 1 \leq i \leq \gamma, 1 \leq j \leq \kappa \} \rightarrow \{0,1,\dots, M-1\},$$
where $\mathcal{C}_{i,j}$ is the circulant corresponding to $1$ at entry $(i,j)$ of the base matrix $\mathbf{{H}^{\textup{g}}}$, and $F(\mathcal{C}_{i,j})$ is the index of the auxiliary matrix to which $\mathcal{C}_{i,j}$ is located. $F^{-1}(0)$ is the set of non-relocated circulants, and the percentage $\mathcal{T}$ of relocated circulants is called the MD density. Conveniently, we define the $\gamma \times \kappa$ matrix $\mathbf{M}$ with $\mathbf{M}(i,j)=F(\mathcal{C}_{i,j})$ as the relocation matrix. An MD-SC code is uniquely determined by the matrices $\mathbf{K}$, $\mathbf{L}$, and $\mathbf{M}$, and the parameters of an MD-SC code are described by the tuple $(\gamma, \kappa, z, L, m, M)$. 
\par
A cycle of length $2g$ (cycle-$2g$) in the Tanner graph of $\mathbf{H}_{\textup{MD}}$ is a $2g$-tuple $(i_1,j_1,i_2,j_2,\dots,i_g,j_g)$, where $(i_k,j_k)$ and $(i_k,j_{k+1})$, $1 \leq k \leq g, \textup{ } j_{g+1}=j_1$, are the positions of nonzero (NZ) entries in $\mathbf{H}_{\textup{MD}}$. Short cycles are detrimental to the performance of an LDPC code as they result in dependencies in decoding and they combine to create absorbing sets \cite{absorbing, channel_aware}. We focus on cycles-$k$, $k \in \{6,8\}$, in this paper, i.e., relocations are done in order to eliminate such cycles when the underlying SC code has girth, i.e., shortest cycle length, $6$ or $8$. The necessary and sufficient condition for a cycle-$2g$ in $\mathbf{H}_{\textup{SC}}$ to create cycles-$2g$ in $\mathbf{H}_{\textup{MD}}$ (remains \textit{active}) after relocations is studied in \cite{homa-hareedy}.

\section{Novel Framework for Probabilistic MD-SC Code Design}\label{sec: framework}

In this section, we present our novel probabilistic design framework that searches for a locally-optimal probability distribution for auxiliary matrices of an MD-SC code. We define a joint probability distribution over the component matrix and the auxiliary matrix (including $\mathbf{X}_0$) to which a $1$ in the base matrix is assigned. This distribution is characterized by an $(m+1) \times M$ so-called probability-distribution matrix. We then formulate an optimization problem and find its solution form to obtain the probability-distribution matrix. Finally, we express the expected number of short cycles in $\mathbf{H}_{\textup{MD}}$ as a function of this probability distribution. The matrix $\mathbf{H}_{\textup{MD}}$ will be designed based on this distribution in Section~\ref{sec: algorithms}.

\subsection{Probabilistic Metrics}
In this subsection, we define metrics relating the joint probability distribution to the expected number of cycle candidates in the MD protograph. A cycle candidate is a way of traversing a protograph pattern to reach a cycle after lifting (see \cite{channel_aware}).

\begin{figure*}[ht!] 
\noindent\makebox[\linewidth]{\rule{\textwidth}{1,5pt}} \\
\begin{align} \label{eqn: prob6}
&P_6(\mathbf{p}^{\textup{con}})=\mathbb{P}\left[\sum_{k=1}^{3}[\mathbf{K}(i_{k},j_{k})-\mathbf{K}(i_{k},j_{k+1})]=0,  \sum_{k=1}^{3} [\mathbf{M}({i_k,j_k})-\mathbf{M}({i_k,j_{k+1}})] \equiv 0 \textup{ (mod $M$)} \right] \nonumber \\
=&\sum\limits_{\substack{\sum\nolimits_{k=1}^{3}(x_k-y_k) = 0, \\ \sum\nolimits_{k=1}^{3} (u_k-v_k) \equiv 0 \textup{ (mod $M$)}}} \prod_{k=1}^3\mathbb{P}\left[ \mathbf{K}(i_{k},j_{k})=x_k, \mathbf{M}({i_k,j_k})=u_k, \mathbf{K}(i_{k},j_{k+1})=y_k, \mathbf{M}({i_k,j_{k+1}})=v_k \right] \nonumber \\
=&\sum\limits_{\substack{\sum\nolimits_{k=1}^{3}(x_k-y_k) = 0, \\ \sum\nolimits_{k=1}^{3} (u_k-v_k) \equiv 0 \textup{ (mod $M$)}}} p_{x_1,u_1}p_{x_2,u_2}p_{x_3,u_3}p_{y_1,v_1}p_{y_2,v_2}p_{y_3,v_3} \nonumber \\
=&\sum_{M \vert b} \left[\sum\limits_{\substack{x_k,y_k \in \{0,1, \dots, m\}, \\ u_k,v_k \in \{0,1, \dots, M-1\}}} p_{x_1,u_1}p_{x_2,u_2}p_{x_3,u_3}p_{y_1,v_1}p_{y_2,v_2}p_{y_3,v_3} X^{x_1+x_2+x_3-y_1-y_2-y_3}Y^{u_1+u_2+u_3-v_1-v_2-v_3} \right]_{0,b} \nonumber \\
=&\sum_{M \vert b} \left[f^3(X,Y)f^3(X^{-1},Y^{-1})\right]_{0,b}.\tag{9}
\end{align}
\noindent\makebox[\linewidth]{\rule{\textwidth}{1,5pt}}
\end{figure*}

\begin{definition} Let $m \geq 1$ and $M \geq 2$. Then, 
\begin{align} \mathbf{P} = \begin{bmatrix} p_{0,0} & p_{0,1} & \dots & p_{0,M-1} \\ p_{1,0} & p_{1,1} & \dots & p_{1,M-1} \\ 
\vdots & \vdots & \ddots & \vdots 
\\ p_{m,0} & p_{m,1} & \dots & p_{m,M-1} \end{bmatrix}_{(m+1) \times M},
\end{align}
where 
\begin{equation} \label{constraint_overall} \sum_{i=0}^{m} \sum_{j=0}^{M-1} p_{i,j} = 1 
\end{equation}
and each $p_{i,j} \in [0,1]$ specifies the probability that a $1$ in $\mathbf{H}_{\textup{MD}}^{\textup{g}}$ is assigned to the $i^{\textup{th}}$ component of the $j^{\textup{th}}$ auxiliary matrix. Thus, $\mathbf{P}$ is referred to as the \textbf{(joint) probability-distribution matrix}. Recall that $\mathbf{X}_0 \triangleq \mathbf{H}_{\textup{SC}}'$. The two-variable \textbf{coupling polynomial} of an MD-SC code associated with the probability-distribution matrix $\mathbf{P}$ is
\begin{equation}
f(X,Y;\mathbf{P}) \triangleq \sum_{i=0}^{m} \sum_{j=0}^{M-1} p_{i,j}X^i Y^j,
\end{equation}
which is abbreviated as $f(X,Y)$ when the context is clear.

\end{definition}
\begin{notation} \label{notation: a-con} For a matrix $\mathbf{A}$ of size $k \times \ell$, the vector $\mathbf{a}^{\textup{con}}=[a_{r}]_{1 \times k\ell}$ is defined as the vector obtained by concatenating the rows of the matrix $\mathbf{A}_{k \times \ell}$ from top to bottom, i.e., $a_{\ell i+j}$ is the entry of $\mathbf{A}$ at $(i,j)$ for $0 \leq i \leq k-1$ and $0 \leq j \leq \ell-1$.
\end{notation}

\begin{definition} The vector $\mathbf{p}^{\textup{con}}$ corresponding to the probability-distribution matrix $\mathbf{P}$ is called the \textbf{probability-distribution vector}. 
\end{definition}

\begin{remark} \label{rmk: order} In the construction of $\mathbf{H}_{\textup{MD}}$ via FL-AO Algorithm in Section \ref{sec: algorithms}, relocations are performed after partitioning and lifting. The respective order in the set-up of Theorems \ref{thm: expected number of cycle-6}, \ref{thm: expected number of cycle-8}, and \ref{thm: forecast} below, however, is partitioning, relocations, then lifting. This change in order, of course, does not affect $\mathbf{H}_{\textup{MD}}$ nor the (expected) cycle counts.
\end{remark}

\begin{theorem} \label{thm: expected number of cycle-6} Let $[\cdot]_{i,j}$ denote the coefficient of $X^iY^j$ in a two-variable polynomial. Denote by $P_6(\mathbf{p}^{\textup{con}})$ the probability of a cycle-$6$ candidate in the base matrix becoming a cycle-$6$ candidate in the MD protograph $\mathbf{H}_{\textup{MD}}^{\textup{g}}$ under random partitioning and relocations with respect to the probability-distribution vector $\mathbf{p}^{\textup{con}}$. Then, we have
\vspace{-0.1em}\begin{equation} \label{exp: prob6} P_6(\mathbf{p}^{\textup{con}}) = \sum_{M \vert b}  \left[ f^3(X,Y)f^3(X^{-1},Y^{-1}) \right]_{0,b}.
\end{equation}
Thus, the expected number $N_6(\mathbf{p}^{\textup{con}})$ of cycle-$6$ candidates in  $\mathbf{H}_{\textup{MD}}^{\textup{g}}$ is given by
\begin{equation} \label{eqn: expected number of cycle-6} N_6(\mathbf{p}^{\textup{con}})= 6 \binom{\gamma}{3} \binom{\kappa}{3} \sum_{M \vert b}  \left[ f^3(X,Y)f^3(X^{-1},Y^{-1}) \right]_{0,b}.
\end{equation}\setcounter{equation}{9}\vspace{-1.3em}
\begin{proof} 
Denote a cycle-$6$ candidate $C$ in the base matrix $\mathbf{{H}^{\textup{g}}}$ by $(i_1,j_1,i_2,j_2,i_3,j_3)$, where $(i_{k},j_{k})$ and $(i_{k},j_{k+1})$, $1\leq k\leq 3$, $j_{4}=j_1$, are edges of $C$ in $\mathbf{{H}^{\textup{g}}}$. From the partitioning condition in \cite{battaglioni} and the relocation condition in \cite{homa-hareedy}, $P_6(\mathbf{p}^{\textup{con}})$ is given by the joint probability of both condition equations being satisfied, and this probability is derived in (\ref{eqn: prob6}). Note that the number of cycle-$6$ candidates in $\mathbf{{H}^{\textup{g}}}$ is $6\binom{\gamma}{3} \binom{\kappa}{3}$ and each becomes a cycle-$6$ candidate in $\mathbf{H}_{\textup{MD}}^{\textup{g}}$ with probability $P_6(\mathbf{p}^{\textup{con}})$. Therefore, $N_6(\mathbf{p}^{\textup{con}})= 6\binom{\gamma}{3} \binom{\kappa}{3} P_6(\mathbf{p}^{\textup{con}})$.
\end{proof}
\end{theorem}

\begin{theorem} \label{thm: expected number of cycle-8}
Denote by $N_8(\mathbf{p}^{\textup{con}})$ the expected number of cycle-$8$ candidates (see Fig.~\ref{fig: cycle8_patterns}) in the MD protograph. Then,
\begin{align}\label{eqn: expected number of cycle-8}
&N_8(\mathbf{p}^{\textup{con}}) \nonumber \\
&= \sum_{M \vert b} \Big\{ w_1 \left[ f^2(X^2,Y^2)f^2(X^{-2},Y^{-2}) \right]_{0,b}  \nonumber \\
&+ w_2 \left[ f(X^2,Y^2)f(X^{-2},Y^{-2})f^2(X,Y)f^2(X^{-1},Y^{-1}) \right]_{0,b} \nonumber \\
&+ w_3  \left[ f(X^2,Y^2)f^2(X,Y)f^4(X^{-1},Y^{-1}) \right]_{0,b} \nonumber \\
&+  w_4  \left[ f^4(X,Y)f^4(X^{-1},Y^{-1}) \right]_{0,b} \Big\},
\end{align}
where $w_1 = \binom{\gamma}{2} \binom{\kappa}{2}$, $w_2 = 3\binom{\gamma}{2} \binom{\kappa}{3}+ 3\binom{\gamma}{3} \binom{\kappa}{2}$, $w_3= 18 \binom{\gamma}{3} \binom{\kappa}{3}$, $ w_4 = 6 \binom{\gamma}{2} \binom{\kappa}{4}+6\binom{\gamma}{4} \binom{\kappa}{2} +36 \binom{\gamma}{3} \binom{\kappa}{4}+36\binom{\gamma}{4} \binom{\kappa}{3}+72 \binom{\gamma}{4} \binom{\kappa}{4}$ if $\gamma \geq 4$, and $w_4 = 6 \binom{\gamma}{2} \binom{\kappa}{4}+36 \binom{\gamma}{3} \binom{\kappa}{4}$ if $\gamma = 3$, where $\kappa \geq 4$.

\begin{proof}
Observe that there are nine protograph patterns that can produce cycles-$8$ after lifting \cite{channel_aware} (see Fig.~\ref{fig: cycle8_patterns}). Following the logic in Theorem \ref{thm: expected number of cycle-6}, \cite[Theorem~1]{GRADE}, and their proofs, we can derive the result above (see also \cite[Chapter~5]{ahh_phd}).
\end{proof}
\end{theorem}

\begin{remark} In this paper, we do not discuss cycles-$4$ in the analysis since removing all of them via informed lifting is easy. We also consider prime $z$, which implies that protograph cycles-$4$ cannot produce cycles-$8$ after lifting. Thus, and since relocations come after lifting in the FL code construction, we take $w_1=0$ in the version of Algorithm~\ref{algo: MD-GRADE6} that handles cycles-$8$ in Section \ref{sec: algorithms} to obtain the probability-distribution matrix.
\end{remark}

\subsection{Gradient-Descent MD-SC Solution Form} 

In this section, we derive the gradient-descent MD-SC solution form by analyzing the Lagrangian of the objective functions presented above. This solution form is a key ingredient in the MD gradient-descent (MD-GRADE) algorithm in Section \ref{sec: algorithms}. Our algorithm produces near-optimal relocation percentages for gradient-descent MD-SC (GD-MD) codes with arbitrary memory of the underlying SC code and arbitrary number of auxiliary matrices.

\begin{remark} The vector $\mathbf{p}^*$ of the probabilities $p_i^*$'s in the constraints of Lemma~\ref{lemma: lagrange6} and Lemma~\ref{lemma: lagrange8} introduced below are set to a locally-optimal edge distribution for the underlying SC code obtained as in \cite{GRADE}. Since they add up to $1$, ($\ref{constraint_overall}$) is automatically satisfied. We chose this approach in order to apply MD relocations to the best underlying SC code. These probabilities are used as an input to Algorithm~\ref{algo: MD-GRADE6} in Section~\ref{sec: algorithms}.
\end{remark} 

\begin{lemma} \label{lemma: lagrange6} For $P_6(\mathbf{p}^{\textup{con}})$ to be locally minimized subject to the constraints 
\begin{equation} 
\begin{split} & p_{i,0}+p_{i,1} + \dots + p_{i,M-1} = p_i^*,
\end{split}
\end{equation}
for all $i \in \{0,1,\dots,m\}$, where each $p_{i,j} \in [0,1]$, it is necessary that the following equations hold for some $c_i \in \mathbb{R}$:
\begin{equation} \label{eq_6}
\begin{split} &\sum_{M \vert b} \left[ f^3(X,Y)f^2(X^{-1},Y^{-1}) \right]_{i,b+j} = c_i, \\
\end{split}
\end{equation}
for all $i \in \{0,1,\dots, m\}$ and $j \in \{0,1, \dots, M-1\}$.

\begin{proof} Since the constraints of the optimization problem in hand satisfy the LICQ regularity property, Karush-Kuhn-Tucker (KKT) conditions must hold. Note that the vector $\mathbf{{r}}^{\textup{con}}$ to appear below is obtained by concatenating the rows of the $(m+1) \times M$ matrix 
\begin{equation} 
\mathbf{R} = \begin{bmatrix} r_0 & r_0 & \dots & r_0 \\ r_1 & r_1 & \dots & r_1 \\ 
\vdots & \vdots & \ddots & \vdots 
\\ r_m & r_m & \dots & r_m \end{bmatrix}.
\end{equation}
We now consider the Lagrangian $L_6(\mathbf{p}^{\textup{con}}) = P_6(\mathbf{p}^{\textup{con}})+ \sum_{i=0}^m r_i (p_i^*-p_{i,0}-p_{i,1} - \dots - p_{i,M-1})$ and compute its gradient as follows:
\begin{align} &\nabla_{\mathbf{p}^{\textup{con}}}  L_6(\mathbf{p}^{\textup{con}}) \nonumber \\
&=  \nabla_{\mathbf{p}^{\textup{con}}} (P_6(\mathbf{p}^{\textup{con}})+ \sum_{i=0}^m r_i (p_i^*-p_{i,0}-p_{i,1} - \dots - p_{i,M-1})) \nonumber \\ 
&= \nabla_{\mathbf{p}^{\textup{con}}} \left( \sum_{M \vert b} \left[ f^3(X,Y)f^3(X^{-1},Y^{-1}) \right]_{0,b} \right) - \mathbf{{r}}^{\textup{con}} \nonumber \\
&= \sum_{M \vert b} \left[ \nabla_{\mathbf{p}^{\textup{con}}}  (f^3(X,Y)f^3(X^{-1},Y^{-1}))\right]_{0,b} - \mathbf{r}^{\textup{con}} \nonumber \\
&= \sum_{M \vert b} \Big\{ 3 \left[f^3(X,Y)f^2(X^{-1},Y^{-1})\nabla_{\mathbf{p}^{\textup{con}}}f(X^{-1},Y^{-1}) \right]_{0,b} \nonumber \\ 
&+  3 \left[ f^2(X,Y)f^3(X^{-1},Y^{-1})\nabla_{\mathbf{p}^{\textup{con}}}f(X,Y) \right]_{0,b} \Big\} - \mathbf{r}^{\textup{con}} \nonumber \\
&= 6  \sum_{M \vert b}  \left[ f^3(X,Y)f^2(X^{-1},Y^{-1})\mathbf{v}_1^{\textup{con}}) \right]_{0,b} - \mathbf{r}^{\textup{con}}, 
\end{align} 
where the last equality follows from the polynomial-symmetry observation $\left[ f(X,Y)\nabla_{\mathbf{p}^{\textup{con}}}f(X^{-1},Y^{-1}) \right]_{0,b} = \left[ f(X^{-1},Y^{-1})\nabla_{\mathbf{p}^{\textup{con}}}f(X,Y) \right]_{0,b}$. Here, $\mathbf{v}_k^{\textup{con}}$ is the vector of length $(m+1)M$ obtained by concatenating the rows of the following $(m+1) \times M$ matrix (of monomials)
\begin{equation} \mathbf{V}_k = \begin{bmatrix} 1 & Y^{-k} & \dots & Y^{-k(M-1)} \\ X^{-k} & X^{-k}Y^{-k} & \dots & X^{-k}Y^{-k(M-1)} \\ 
\vdots & \vdots & \ddots & \vdots 
\\ X^{-km} & X^{-km}Y^{-k} & \dots & X^{-km}Y^{-k(M-1)} \end{bmatrix}
\end{equation}
for $k \in \mathbb{Z}$. When $P_6(\mathbf{p}^{\textup{con}})$ reaches its local minimum, $\nabla_{\mathbf{p}^{\textup{con}}}  L_6(\mathbf{p}^{\textup{con}}) = \mathbf{0}_{1\times (m+1)M},$
which directly leads to (\ref{eq_6}) by defining $c_i = r_i /6$ for all $i \in \{0,1,\dots,m\}$.
\end{proof}
\end{lemma}

\begin{lemma} \label{lemma: lagrange8} For $N_8(\mathbf{p}^{\textup{con}})$ to be locally minimized subject to the constraints
\begin{equation} 
\begin{split} & p_{i,0}+p_{i,1} + \dots + p_{i,M-1} = p_i^*,
\end{split}
\end{equation}
for all $i \in \{0,1,\dots,m\}$, where each $p_{i,j} \in [0,1]$, it is necessary that the following equations hold for some $c_i \in \mathbb{R}$:
\begin{align} \label{eq_8}
&\sum_{M \vert b} \Big\{ w_1 \left[ 4f^2(X^2,Y^2)f(X^{-2},Y^{-2}) \right]_{2i,b+2j} \nonumber \\
&+ w_2 \left[ 2f(X^2,Y^2)f^2(X,Y)f^2(X^{-1},Y^{-1}) \right]_{2i,b+2j} \nonumber \\
&+ w_2 \left[ 4f(X^2,Y^2)f(X^{-2}Y^{-2})f^2(X,Y)f(X^{-1},Y^{-1}) \right]_{i,b+j} \nonumber \\
&+ w_3 \left[ f^2(X,Y)f^4(X^{-1},Y^{-1}) \right]_{-2i,b-2j} \nonumber \\ 
&+ w_3 \left[ 2f(X^2,Y^2)f(X,Y)f^4(X^{-1},Y^{-1}) \right]_{-i,b-j} \nonumber \\  
&+ w_3 \left[ 4f(X^2,Y^2)f^2(X,Y)f^3(X^{-1},Y^{-1}) \right]_{i,b+j} \nonumber \\ 
&+  w_4 \left[ 8f^4(X,Y)f^3(X^{-1},Y^{-1}) \right]_{i,b+j} \Big\} \nonumber \\
&= c_i,
\end{align}
for all $i \in \{0,1, \dots, m\}$ and $j \in \{0,1, \dots, M-1\}$.

\begin{proof}
We consider the Lagrangian $L_8(\mathbf{p}^{\textup{con}}) = N_8(\mathbf{p}^{\textup{con}})+ \sum_{i=0}^m r_i (p_i^*-p_{i,0}-p_{i,1} - \dots - p_{i,M-1})$ and compute its gradient as follows:
\begin{align}\label{eqn: cycle_8_grad}
&\nabla_{\mathbf{p}^{\textup{con}}}  L_8(\mathbf{p}^{\textup{con}}) \nonumber \\
&=  \nabla_{\mathbf{p}^{\textup{con}}} (N_8(\mathbf{p}^{\textup{con}})+ \sum_{i=0}^m r_i (p_i^*-p_{i,0}-p_{i,1} - \dots - p_{i,M-1})) \nonumber \\ 
&= \sum_{M \vert b} \nabla_{\mathbf{p}^{\textup{con}}} \Big\{ w_1 \left[ f^2(X^2,Y^2)f^2(X^{-2},Y^{-2}) \right]_{0,b} \nonumber \\
&+ w_2 \left[ f(X^2,Y^2)f(X^{-2},Y^{-2})f^2(X,Y)f^2(X^{-1},Y^{-1}) \right]_{0,b} \nonumber \\
&+ w_3 \left[ f(X^2,Y^2)f^2(X,Y)f^4(X^{-1},Y^{-1}) \right]_{0,b} \nonumber \\
&+  w_4 \left[ f^4(X,Y)f^4(X^{-1},Y^{-1}) \right]_{0,b} \Big\} - \mathbf{r}^{\textup{con}} \nonumber \\
&= \sum_{M \vert b} \Big\{ w_1 \left[ 4f^2(X^2,Y^2)f(X^{-2},Y^{-2})\mathbf{v}_2^{\textup{con}} \right]_{0,b} \nonumber \\
&+ w_2 \left[ 2f(X^2,Y^2)f^2(X,Y)f^2(X^{-1},Y^{-1})\mathbf{v}_2^{\textup{con}}]_{0,b} \right. \nonumber \\
&+ w_2 \left[ 4f(X^2,Y^2)f(X^{-2}\hspace{-0.3em},Y^{-2})f^2(X,Y)f(X^{-1}\hspace{-0.3em},Y^{-1})\mathbf{v}_1^{\textup{con}} \right]_{0,b} \nonumber \\
&+ w_3 \left[ f^2(X,Y)f^4(X^{-1},Y^{-1})\mathbf{v}_{-2}^{\textup{con}} \right]_{0,b} \nonumber \\ 
&+ w_3 \left[ 2f(X^2,Y^2)f(X,Y)f^4(X^{-1},Y^{-1}) \mathbf{v}_{-1}^{\textup{con}} \right]_{0,b} \nonumber \\  
&+ w_3 \left[ 4f(X^2,Y^2)f^2(X,Y)f^3(X^{-1},Y^{-1}) \mathbf{v}_1^{\textup{con}} \right]_{0,b} \nonumber \\ 
&+  w_4 \left[ 8f^4(X,Y)f^3(X^{-1},Y^{-1})\mathbf{v}_1^{\textup{con}} \right]_{0,b} \Big\} - \mathbf{r}^{\textup{con}}.
\end{align}
When $N_8(\mathbf{p}^{\textup{con}})$ reaches its local minimum, $\nabla_{\mathbf{p}^{\textup{con}}}  L_8(\mathbf{p}^{\textup{con}}) = \mathbf{0}_{1 \times (m+1)M},$ which directly leads to (\ref{eq_8}) by defining $c_i = r_i$ for all $i \in \{0,1,\dots,m\}$.
\end{proof}
\end{lemma}

\subsection{Outcomes Forecasted} \label{sec: forecast}

In this section, we compute the expected number of cycles-$k$, $k \in \{6,8\}$, under a specific MD probability distribution (which gives an estimate of what the FL-AO algorithm can produce; see Section~\ref{sec: design}). These numbers will inform us what to expect from incorporating the probability-distribution matrix in designing GD-MD codes under random partitioning and lifting. The GD-MD codes we design will have lower numbers of cycles (of length $=$ the code girth) than the upper bound of the expected number, where partitioning and lifting matrices are designed as in \cite{GRADE} and \cite{oocpo}.

\begin{figure}
\centering
\includegraphics[width=0.42\textwidth]{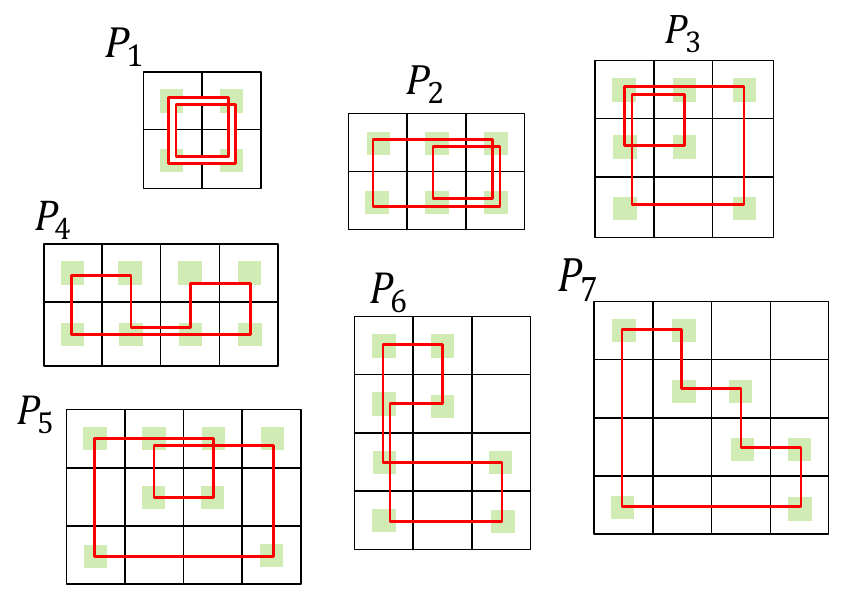}
\caption{Seven (out of nine) protograph patterns that can result in cycles-$8$ in the Tanner graph after lifting. The transposes of $P_2$ and $P_4$ are excluded from the list. Light green squares are the non-zero entries specifying the cycle-candidate edges.}
\label{fig: cycle8_patterns}
\end{figure}

\begin{theorem} \label{thm: forecast} After random partitioning, relocations, and lifting based on a probability-distribution vector $\mathbf{p}^{\textup{con}}$, the expected number $E_6(\mathbf{p}^{\textup{con}})$ of cycles-$6$ in the Tanner graph of $\mathbf{H}_{\textup{MD}}$ is
\begin{equation} \label{Eobj6mid} E_6(\mathbf{p}^{\textup{con}}) \approx N_6(\mathbf{p}^{\textup{con}})\cdot \frac{(2L-m)}{2}\cdot M,
\end{equation}
and the expected number $E_8(\mathbf{p}^{\textup{con}})$ of cycles-$8$ is 
\begin{equation} \label{Eobj8mid}
E_8(\mathbf{p}^{\textup{con}}) \approx N_8(\mathbf{p}^{\textup{con}}) \cdot (L-m) \cdot M,
\end{equation}
where $L > \chi$ and $\chi$ is the maximum number of consecutive SC replicas a cycle of interest can span ($\kappa \geq 4$).

\begin{proof}
Let $E_{\textup{obj}}$ be the expected total number of cycle-$k$ candidates out of all cycle-$k$ candidates $\{C_i \,|\, i \in \mathcal{I}\}$ in the all-one base matrix $\mathbf{{H}^{\textup{g}}}$ that remain active after random partitioning, random relocations, and random lifting (see Remark \ref{rmk: order}). Here, $\mathcal{I}$ is the set of candidate indices. Define $\mathbbm{1}(C_i)$ as an indicator function on $C_i$ being active. Then,
\begin{align} 
E_{\textup{obj}} &=\mathbb{E} \bigg[ \sum_{i \in \mathcal{I}} \mathbbm{1}(C_i) \bigg] \nonumber \\
&= \sum_{i \in \mathcal{I}}  \mathbb{E} [ \mathbbm{1}(C_i) ] \nonumber \\
&= \sum_{P \in \mathcal{K}(C)} \sum_{i \in \mathcal{I}(P)}  \mathbb{E}_{P} [ \mathbbm{1}(C_i) ] ,
\end{align}
where $\mathcal{K}(C)$ is the set of all protograph patterns that can produce the cycle-$k$ of interest denoted by $C$, $\mathcal{I}(P)$ is the set of indices of all cycle candidates in $\mathbf{{H}^{\textup{g}}}$ associated with pattern $P$. Moreover,
\begin{equation}
\mathbb{E}_{P} [ \mathbbm{1}(C_i) ]= \mathbb{P}(K^{C_i}_{P}, F^{C_i}_{P}) \mathbb{P}(L^{C_i}_{P} \mid  K^{C_i}_{P}, F^{C_i}_{P}),
\end{equation}
where $\mathbb{P}(K^{C_i}_{P}, F^{C_i}_{P})$ is the probability that $C_i$ remains active after partitioning and relocations, while $\mathbb{P}(L^{C_i}_{P} \mid  K^{C_i}_{P}, F^{C_i}_{P})$ is the probability that $C_i$ remains active after lifting given that it arrives at this stage active ($K^{C_i}_{P}$, $F^{C_i}_{P}$, and $L^{C_i}_{P}$ are the associated events). Observe that what happens at the lifting stage depends on partitioning and relocations.
\par 
First, note that there is only one pattern $P$ (a cycle-$6$ itself) in the base matrix that can produce cycles-$6$ in $\mathbf{H}_{\textup{MD}}$, i.e., $|\mathcal{K}(C)|=1$. Furthermore,
\begin{enumerate} 
\item[a)] $\mathbb{P}(K^{C_i}_{P}, F^{C_i}_{P})=P_6(\mathbf{p}^{\textup{con}})$, 
\item[b)] $|\mathcal{I}|= 6 \binom{\gamma}{3} \binom{\kappa}{3}$, and
\item[c)] $\mathbb{P}(L^{C_i}_{P} \mid  K^{C_i}_{P}, F^{C_i}_{P})=1/z$. 
\end{enumerate}
Hence and using (\ref{eqn: expected number of cycle-6}), $ E_{\textup{obj}}$ is approximated by $N_6(\mathbf{p}^{\textup{con}})/z$. Combining this with the fact that an active candidate $C_i$ produces multiple copies after each stage (specifically $z$ after lifting), we obtain 
\begin{align} N_6(\mathbf{p}^{\textup{con}})&\cdot (L-s_{\textup{max}}+1)\cdot M \nonumber  \\
& \leq E_6(\mathbf{p}^{\textup{con}}) \nonumber \\
& \leq N_6(\mathbf{p}^{\textup{con}})\cdot (L-s_{\textup{min}}+1)\cdot M,
\end{align}
where $s_{\textup{min}}$ and $s_{\textup{max}}$ are the minimum and maximum spans, in terms of replicas, of a cycle-$6$ in $\mathbf{H}_{\textup{SC}}$. Note that $s_{\textup{min}}=1$ and $s_{\textup{max}}=\chi=m+1$. By assumption, $L\gg m$, and thus we can take $E_6(\mathbf{p}^{\textup{con}}) \approx E_{\textup{obj}}\cdot \frac{(2L-m)}{2}\cdot M$ via averaging $L$ and $L-m$ as an estimate, which gives (\ref{Eobj6mid}).
\par 
For the second part of the statement, observe that there are nine patterns in the base matrix that can produce cycles-$8$ in $\mathbf{H}_{\textup{MD}}$ \cite{GRADE, channel_aware}, i.e., $|\mathcal{K}(C)|=9$. We will find the expected number of cycles-$8$ in $\mathbf{H}_{\textup{MD}}$ for pattern $P_4$ shown in Fig.~\ref{fig: cycle8_patterns} as this is a quite rich case.

We first note that
\begin{enumerate} 
\item[a)] $\mathbb{P}(K^{C_i}_{P_4}, F^{C_i}_{P_4})= \sum_{M \vert b} [ f^4(X,Y)f^4(X^{-1},Y^{-1})]_{0,b}$ and
\item[b)] $|\mathcal{I}|= 6\binom{\gamma}{2} \binom{\kappa}{4}$.
\end{enumerate}
Next, we show that $(z-1)(z-2)/z^3 \leq \mathbb{P}(L^{C_i}_{P_4} \mid  K^{C_i}_{P_4}, F^{C_i}_{P_4}) \leq 1/z$ for lifting.
\par
Note that $P_4$ is formed via three basic (fundamental) cycles-$4$. Following the arguments of \cite{fossorier} and \cite[Theorem~3]{rohith2}, we consider the following cases for $P_4$ after partitioning:
\begin{enumerate} 
\item[{2.a)}] The three basic cycles-$4$ are kept.
\item[{2.b)}] Only two basic cycles-$4$ are kept. 
\item[{2.c)}] No cycles-$4$ are kept.
\end{enumerate}
Suppose that the $4$ edges of $P_4$ on the first row are labeled N1, N2, N3, N4, starting from the left, and the $4$ edges on the second row are labeled N5, N6, N7, N8, starting from the left. We access these edges in the following order N1, N5, N6, N2, N3, N7, N8, N4.
\par
In Case 2.a), suppose we start from the edge N1, and we trace the possible lifting choices for edges N2, N7, and N4 at the end of each basic cycle traversal. N2 has $z-1$ options for a power of $\sigma$ in order not to form a cycle-$4$ on the left. N7 has $z-2$ options in order not to form a cycle-$4$ from the disjunctive union of the two left-most basic cycles. Finally, N4 has a single option to close the walk and form a cycle-$8$ \cite{fossorier}. Hence, $\mathbb{P}(L^{C_i}_{P_4} \mid  K^{C_i}_{P_4}, F^{C_i}_{P_4})= (z-1)(z-2)(1)/z^3=(z-1)(z-2)/z^3$.
\par
In Case 2.b), suppose we start from the edge N1, and we trace the possible lifting choices for edges N2, N7, and N4 at the end of each basic cycle traversal. N2 has $z-1$ options for a power of $\sigma$ in order not to form a cycle-$4$ on the left. N7 is free to receive any of the $z$ options. Finally, N4 has a single option to close the walk and form a cycle-$8$ \cite{fossorier}. Hence, $\mathbb{P}(L^{C_i}_{P_4} \mid  K^{C_i}_{P_4}, F^{C_i}_{P_4})= (z-1)z(1)/z^3=(z-1)/z^2$. 
\par
In Case 2.c), it is clear that $\mathbb{P}(L^{C_i}_{P_4} \mid  K^{C_i}_{P_4}, F^{C_i}_{P_4})=1/z$. 
\par
Note that $\mathbb{P}(L^{C_i}_{P_4} \mid  K^{C_i}_{P_4}, F^{C_i}_{P_4})$ is closer to $1/z$ because it is more likely to have no cycles-$4$, for $m > 1$, after random partitioning (in which case $C_i$ becomes a true cycle-$8$ in the MD protograph). Therefore, we take $\mathbb{P}(L^{C_i}_{P_4} \mid  K^{C_i}_{P_4}, F^{C_i}_{P_4})=1/z$ as an estimate. The other patterns can be treated in a similar fashion, and thus we take $\mathbb{P}(L^{C_i}_{P} \mid  K^{C_i}_{P}, F^{C_i}_{P})=1/z$ for all possible cycle-$8$ protograph patterns. Hence, $ E_{\textup{obj}}$ is approximated by $N_8(\mathbf{p}^{\textup{con}})/z$. Consequently, we obtain 
\begin{align} N_8(\mathbf{p}^{\textup{con}})&\cdot (L-s_{\textup{max}}+1)\cdot M \nonumber  \\
& \leq E_8(\mathbf{p}^{\textup{con}}) \nonumber \\
& \leq N_8(\mathbf{p}^{\textup{con}})\cdot (L-s_{\textup{min}}+1)\cdot M,
\end{align}
where $s_{\textup{min}}$ and $s_{\textup{max}}$ are the minimum and maximum spans of a cycle-$8$ in $\mathbf{H}_{\textup{SC}}$. Note that $s_{\textup{min}}=1$ and $s_{\textup{max}}=\chi=2m+1$ due to the existence of structures $P_6$ and $P_7$ (only $P_7$) in case $\gamma \geq 4$ ($\gamma = 3$). By assumption, $L\gg m$, and thus we can take $E_8(\mathbf{p}^{\textup{con}}) \approx N_8(\mathbf{p}^{\textup{con}})\cdot (L-m)\cdot M$ via averaging $L$ and $L-2m$ as an estimate, which gives (\ref{Eobj8mid}).
\end{proof}
\end{theorem}

\begin{remark}\label{rmk4} Note that the upper bound of $E_8(\mathbf{p}^{\textup{con}})$ in the proof above is exact. One obtains an exact lower bound as well by studying the probabilities $\mathbb{P}(L^{C_i}_{P} \mid  K^{C_i}_{P}, F^{C_i}_{P})$ for each cycle-$8$ pattern, and it is in fact lower than $N_8(\mathbf{p}^{\textup{con}}) \cdot (L-2m) \cdot M$. However, the expected number $E_8(\mathbf{p}^{\textup{con}})$ is notably closer to the upper bound.
\end{remark}

We now discuss the value of Theorem~\ref{thm: forecast} to the coding theory community.
\begin{enumerate}
\item The GD-MD codes we design will have lower numbers of cycles (of length $=$ girth) than the upper bound of the expected number when the partitioning and lifting matrices are designed as in \cite{GRADE} and \cite{oocpo}. This is expected to hold in general when we use optimized partitioning and lifting matrices designed for the underlying SC~codes. As we shall see, our estimate can get close to the actual finite-length cycle counts.
\item The expected number of cycles-$k$, $k \in \{6,8\}$, in $\mathbf{H}_{\textup{MD}}$ decreases with the relocation percentage $\mathcal{T}$. However, the rate of reduction becomes close to $0$ after some $\mathcal{T}$ value. Hence, one can locate a threshold vector $\mathbf{p}_{\textup{thold}}$ where $E_k(\mathbf{p}_{\textup{thold}})$ is at most, say, $5\%$ more than the minimum $E_k(\mathbf{p}^{\textup{con}}).$ This allows us to specify an informed MD density $\mathcal{T}$ for the FL-AO algorithm (within the margin governed by decoding latency) to potentially decrease its computational complexity.  
\item $E_{\textup{obj}}$ in Theorem~\ref{thm: forecast} gives the expected number of cycles-$k$ produced from the cycle-$k$ candidates in the all-one base matrix $\mathbf{{H}^{\textup{g}}}$ that remain active after random partitioning, random relocation, and random lifting. In case this number is single-digit (at $\mathbf{p}_{\textup{thold}}$ for example), it is likely that the FL-AO algorithm will eliminate all active cycles and produce an MD-SC code with girth $\geq k+2$. Provided that cycles (of length $=$ girth) can in fact be entirely eliminated, $E_{\textup{obj}}$ gives us a threshold for the MD density to achieve this goal.
\item Having the MD-SC degrees of freedom $m$, $L$, and $M$, we can choose these parameters to achieve best performance subject to the design constraints such as the length and rate. For example, if one has an allowed range for the length of MD-SC code to design while the rate is fixed, $M$ can be determined so that $E_{\textup{obj}}$ is as low as possible while the length is within the allowed range. This, of course, requires the code designer to run the MD-GRADE algorithm multiple times (see the next section).
\end{enumerate}

\section{Design and Optimization Algorithms} \label{sec: algorithms} 
\subsection{Multi-Dimensional Gradient-Descent Distributor}

\begin{algorithm}
\caption{Multi-Dimensional Gradient-Descent Distributor (MD-GRADE) for Cycle-$6$ Optimization} \label{algo: MD-GRADE6}
\begin{algorithmic}[1]
\Statex \textbf{Inputs:} $\gamma,\kappa, L, m, M$: parameters of the MD code, $\mathbf{p}^{*}$: locally-optimal edge distribution of an SC code with $\gamma,\kappa,m$ obtained by \cite[Algorithm~1]{GRADE}, $\mathcal{T}$: MD density, and $\epsilon, \alpha$: accuracy and step size of gradient descent.
\Statex \textbf{Outputs:} $\mathbf{P}$: a (joint) locally-optimal probability-distribution matrix,  $E_6$: the expected number of cycles-$6$ in the Tanner graph of $\mathbf{H}_{\textup{MD}}$ obtained by Theorem \ref{thm: forecast}.
\Statex \textbf{Intermediate Variables:} $F_{\textup{prev}}, F_{\textup{cur}}$: the values of the objective function in (\ref{eqn: expected number of cycle-6}) at the previous and current iterations, $\mathbf{G}$: the gradient matrix (of size $(m+1) \times M$) of the objective function where $\mathbf{g}^{\textup{con}}=\nabla_{\mathbf{p}^{\textup{con}}} N_6(\mathbf{p}^{\textup{con}})$, $\mathrm{pct}$: the current percentage of the sum of probabilities in the first column of $\mathbf{P}$.
\State Set $\mathbf{P}$  to the zero matrix $\mathbf{0}_{(m+1) \times M}$, and let $\mathbf{P}[i][0] \gets \mathbf{p}^*[i]$, $\forall i\in \{0,1,\dots,m\}$, $F_{\textup{prev}}=0$.
\State $\mathbf{\overline{P}} \gets \mathrm{mirror}(\mathbf{P})$, $F_{\textup{cur}}=0$, $\mathbf{G}=\mathbf{0}_{(m+1)\times M}$. 
\Statex // \textit{Below we use Python's $\mathrm{range}$ function where $\mathrm{range}(a,b)$ returns the set of integers from $a$ to $b-1$.}
\For{$i \textup{ in } \mathrm{range}(0, 6M-5)$}
\If{$(i + 3 - 3M) \equiv 0\,(\textup{mod}\,M)$}
\State $F_{\textup{cur}} \gets F_{\textup{cur}}+  6\binom{\gamma}{3}\binom{\kappa}{3} \cdot \mathrm{conv}(\mathbf{P},\mathbf{P},\mathbf{P}, \mathbf{\overline{P}},\mathbf{\overline{P}},\allowbreak\mathbf{\overline{P}})[3m][i]$.
\EndIf
\EndFor
\For{$ i \textup{ in } \mathrm{range}(0, m+1)$}
\For{$ j \textup{ in } \mathrm{range}(0, M)$}
\For{$ b \textup{ in } \mathrm{range}(2-2M-j,3M-2-j)$}
\If{$b \equiv 0\,(\textup{mod}\,M)$}
\State $\mathbf{G}[i][j] \gets \mathbf{G}[i][j] +  36\binom{\gamma}{3}\binom{\kappa}{3} \cdot \mathrm{conv}(\mathbf{P},\mathbf{P},\allowbreak \mathbf{P}, \mathbf{\overline{P}},\mathbf{\overline{P}})[i+2m][b+j+2M-2]$.
\EndIf
\EndFor
\EndFor
\EndFor
\State $\mathbf{P} \gets \mathbf{P} - \alpha \displaystyle \frac{\mathbf{G}}{\| \mathbf{g}^{\textup{con}} \|}$.
\State $\mathbf{P} \gets \mathrm{force}(\mathbf{P}, \mathbf{p}^*)$, $\mathrm{pct} \gets 100 \sum_{i=0}^m \mathbf{P}[i][0]$.
\If{$(\mathrm{pct} > 100-\mathcal{T})$ and $(|F_{\textup{cur}}-F_{\textup{prev}}| > \epsilon)$} 
\State $F_{\textup{prev}} \gets F_{\textup{cur}}$.
\State \textbf{go to} Step 2.
\EndIf
\State $E_6 \gets \frac{(2L-m)}{2} \cdot M \cdot F_{\textup{cur}}$.
\State \textbf{return} $\mathbf{P}$ and $E_6$.
\end{algorithmic}
\end{algorithm}

Based on (\ref{eqn: expected number of cycle-6}), we develop a gradient-descent algorithm (MD-GRADE) to obtain a locally-optimal probability distribution for MD-SC codes with low number of cycles-$6$. We first introduce some functions below that are essential in the MD-GRADE algorithm.
\begin{enumerate} 
\item $\mathrm{conv}(\cdot)$ is a self-defined recursive function for 2D convolution that takes $t$ matrices $\mathbf{A}_1,\mathbf{A}_2,\dots,\mathbf{A}_t$ with $\textup{dim}(\mathbf{A}_i)=n_i \times m_i$ and produces the matrix $\mathbf{B}=[b_{r,s}]_{n_\textup{c} \times m_\textup{c}}$, where $n_\textup{c} = 1-t+\sum_{i=1}^t n_i$, $m_\textup{c} = 1-t+\sum_{i=1}^t m_i$,
\[ b_{r,s} = \hspace{-2.5em}\sum_{\substack{0 \leq i \leq \min(n_1-1,r) \\ 0 \leq j \leq \min(m_1-1,s)}} \hspace{-2.5em} \mathbf{A}_1[i][j] \cdot \mathrm{conv}(\mathbf{A}_2,\mathbf{A}_3, \dots,\mathbf{A}_t)[r-i][s-j]\]
if $t \geq 2$, and $\mathrm{conv}(\mathbf{A})=\mathbf{A}$ for any matrix $\mathbf{A}$.
\item $\mathrm{mirror}(\cdot)$ is a self-defined function that takes an $n_\textup{a} \times m_\textup{a}$ matrix $\mathbf{A}=[a_{i,j}]_{n_\textup{a} \times m_\textup{a}}$ and outputs the matrix $\mathbf{B}=[b_{i,j}]_{n_\textup{a} \times m_\textup{a}}=[a_{n_\textup{a}-1-i,m_\textup{a}-1-j}]_{n_\textup{a} \times m_\textup{a}}$. We refer to $\mathrm{conv}(\cdot)$ and $\mathrm{mirror}(\cdot)$ as convolution and reverse functions, respectively. 
\item $\mathrm{force(\cdot)}$ is a function specifically defined for $\mathbf{P}$. It takes the inputs $\mathbf{P}$ and $\mathbf{p}^*$. It projects each row of $\mathbf{P}$ indexed by $t$ onto the hyperplane defined by the normal vector $\mathbf{1}_{1\times M}$ and adds the vector $\mathbf{p}^{*}[t]/M \cdot \mathbf{1}_{1\times M}$ via
$$\mathbf{P}[t][k] \gets \mathbf{P}[t][k] + \bigg[ \mathbf{p}^{*}[t]- \sum_{\ell=0}^{M-1} \mathbf{P}[t][\ell] \bigg]/M.$$
Summations are for non-updated values. Then, it corrects its rows by making the negative entries $0$ and by updating the positive entries in order not to change the sum of each row via
\vspace{-0.5em}$$ \mathbf{P}[t][k] \gets \mathbf{P}'[t][k] \cdot \frac{\sum_{\ell=0}^{M-1} \mathbf{P}[t][\ell]}{\sum_{\ell=0}^{M-1} \mathbf{P}'[t][\ell]},$$
where $\mathbf{P}'[t][k] = \begin{cases} 0 & \textup{ if } \mathbf{P}[t][k] < 0, \\ \mathbf{P}[t][k] & \textup{ otherwise.} \end{cases}$
\end{enumerate}
Here, $\mathbf{p}^{*}$ is a locally-optimal edge distribution of an SC code with parameters $(\gamma, \kappa, m)$ obtained by \cite[Algorithm~1]{GRADE}.

\begin{algorithm}
\caption{Multi-Dimensional Gradient-Descent Distributor (MD-GRADE) for Cycle-$8$ Optimization} \label{algo: MD-GRADE8}
\begin{algorithmic}[1]
\State Set $\mathbf{P}$ to the zero matrix $\mathbf{0}_{(m+1) \times M}$, and let $\mathbf{P}[i][0] \gets \mathbf{p}^*[i]$, $\forall i \in \{0,1,\dots,m\}$.
\State $\mathbf{\overline{P}} \gets \mathrm{mirror}(\mathbf{P})$, $\mathbf{P}_2 \gets \mathrm{square}(\mathbf{P})$, $F_{\textup{cur}}=0$, $F_{\textup{cur}}^3=0$, $\mathbf{G}^2=\mathbf{0}_{(m+1)\times M}$, $\dots$.
\State $\dots$
\For{$i \textup{ in } \mathrm{range}(0, 8M-7)$}
\If{$(i + 4 - 4M) \equiv 0\,(\textup{mod}\,M)$}
\State $F_{\textup{cur}}^3 \gets F_{\textup{cur}}^3+ 18 \binom{\gamma}{3} \binom{\kappa}{3} \cdot \mathrm{conv}(\mathbf{P}_2, \mathbf{P},\mathbf{P}, \mathbf{\overline{P}},\mathbf{\overline{P}}, \allowbreak \mathbf{\overline{P}},\mathbf{\overline{P}})[4m][i]$.
\EndIf
\EndFor
\State $\dots$ 
\For{$ i \textup{ in } \mathrm{range}(0, m+1)$}
\For{$ j \textup{ in } \mathrm{range}(0, M)$}
\For{$ b \textup{ in } \mathrm{range}(2-2M-2j,4M-3-2j)$}
\If{$b \equiv 0\,(\textup{mod}\,M)$}
\State  $\mathbf{G}^2[i][j] \gets \mathbf{G}^2[i][j] +  \bigg[ 6 \binom{\gamma}{3} \binom{\kappa}{2} + 6 \binom{\gamma}{2} \binom{\kappa}{3} \bigg] \cdot \mathrm{conv}(\mathbf{P}_2,\mathbf{P},\mathbf{P}, \mathbf{\overline{P}},\mathbf{\overline{P}})[2(i+m)][b+2j+2M-2]$
\EndIf
\EndFor
\EndFor
\EndFor
\State $\dots$ 
\State \textbf{return} $\mathbf{P}$ and $E_8$.
\end{algorithmic}
\end{algorithm}

We need to make the following adjustments in Algorithm~\ref{algo: MD-GRADE6} in order to obtain its version for cycles-$8$:
\begin{enumerate} 
\item We need another function for the role of the factor $f(X^2,Y^2)$ in $N_8(\mathbf{p}^{\textup{con}})$:\\
$\mathrm{square}(\cdot)$ is a self-defined function that takes an $n_\textup{a} \times m_\textup{a}$ matrix $\mathbf{A}=[a_{i,j}]_{n_\textup{a} \times m_\textup{a}}$ and outputs the matrix $\mathbf{B}=[b_{r,s}]_{2n_\textup{a}-1 \times 2m_\textup{a}-1}$ with 
\[ b_{r,s}= \begin{cases} a_{r/2, s/2} & \textup{ if } r \textup{ and } s \textup{ are even,} \\
0 & \textup{ otherwise.}
\end{cases} \]
This function appears in Step~2 to obtain an auxiliary matrix $\mathbf{P}_2$ that will be used in the computations of $F_{\textup{cur}}$ and the gradient matrix $\mathbf{G}$. 
\item The assignment of $F_{\textup{cur}}$ needs to be adjusted based on the expression in (\ref{eqn: expected number of cycle-8}), and the ranges in previous steps need to be altered accordingly.
\item The assignment of $\mathbf{G}$ needs to be adjusted based on the gradient in (\ref{eq_8}) and (\ref{eqn: cycle_8_grad}), and the ranges in previous steps need to be altered accordingly.
\end{enumerate}
\par 
Algorithm~\ref{algo: MD-GRADE8} illustrates the required adjustments above for the sample term $$F_{\textup{cur}}^3=  \sum_{M \vert b} w_3  \left[ f(X^2,Y^2)f^2(X,Y)f^4(X^{-1},Y^{-1}) \right]_{0,b},$$ 
which is the third summand in (\ref{eqn: expected number of cycle-8}) that comes from the contribution of patterns $P_3$, and the term
\begin{align*} & \mathbf{G}^2[i][j]  =  \\
 & \sum_{M \vert b}  \ w_2 \left[ 2f(X^2,Y^2)f^2(X,Y)f^2(X^{-1},Y^{-1}) \right]_{2i,b+2j},
\end{align*}
which is the second summand in (\ref{eq_8}) that comes from the contribution of patterns $P_2$ and its transpose.

\begin{remark} Note that the MD-GRADE algorithm provides us with a non-trivial solution. In particular, the relocation percentages for each component matrix $\mathbf{H}_i$ are not necessarily the same. For instance, for the input parameters $(\gamma,\kappa,L,m,M) = (3,17,100,9,9)$, the vector of component-wise relocation percentages is $[ 14.83 \textup{ } 45.16 \textup{ } 60.75 \textup{ } 66.67 \allowbreak 70.6 \textup{ } 70.6 \textup{ } 66.67 \textup{ } 60.75 \textup{ } 45.16 \textup{ } 14.83],$ where the total relocation percentage is $34.45\%$. This vector shows that more circulants need to be relocated from the middle component matrices of the SC code than the side ones. These percentages are quite close to final relocation percentages produced by the FL-AO to design GD-MD Code~4.1.
\end{remark}

\subsection{Probabilistic Design of GD-MD Codes} \label{sec: design}

We now present the FL-AO algorithm that performs relocations and produces a GD-MD code. This algorithm is initiated with a relocation arrangement based on a probability-distribution matrix specified by the MD-GRADE algorithm.

\begin{algorithm}
\caption{Finite-length Algorithmic Optimizer (FL-AO) for Designing High Performance MD-SC Codes}
\begin{algorithmic}[1]
\Statex \textbf{Inputs:} $\mathbf{H}_{\textup{SC}}$, $M$, an even positive integer $k$, a probability-distribution matrix $\mathbf{P}$, and a relocation bound $(\mathrm{RB})$.
\Statex \textbf{Outputs:} The parity-check matrix $\mathbf{H}_{\textup{MD}}$ of the GD-MD code, the number of cycles-$k$ in $\mathbf{H}_{\textup{MD}}$, and the number of iterations.
\Statex \textbf{Intermediate Variables:} The number $\mathrm{active\_cycle\_cur}$ ($\mathrm{active\_cycle\_prev}$) of active cycles-$k$ in $\mathbf{H}_{\textup{SC}}$ at the current (previous) iteration.
\State Initially, randomly relocate some circulants to auxiliary matrices $\mathbf{X}_1, \dots, \mathbf{X}_{M-1}$ based on $\mathbf{P}$ and update the values $F(\mathcal{C}_{i,j})$, $\forall i,j$, based on relocations. Set $\mathrm{count}=0$, $\mathrm{active\_cycle\_prev} = 0$, and $\mathrm{active\_cycle\_cur}= 0$.
\State Locate all cycles-$k$ in the Tanner graph of $\mathbf{H}_{\textup{SC}}$, and set $\mathrm{active\_cycle\_prev}$ to their number
\State Create a pruned cycle list where each cycle has at least two circulants in the replica $\mathbf{R}_{\lfloor L/2 \rfloor}$.
\State For each circulant, determine the number of active cycles which involve it from the pruned cycle list and select the circulant(s) with the maximum number. If multiple circulants, choose one randomly. Let $\mathcal{C}$ denote the selected circulant.
\State For each $i \in \{0,1,\dots,M-1\}$, determine the number of cycles involving $\mathcal{C}$ that become inactive after relocating $\mathcal{C}$ to $\mathbf{X}_i$, based on the relocation condition in (\ref{eqn: prob6}) (see \cite{homa-hareedy}).
\State Choose the index $i$ associated with the highest number of inactive cycles in Step~5. In case of multiple indices, choose one of them randomly.
\State Relocate the circulant $\mathcal{C}$ from Step~4 to $\mathbf{X}_i$, where $i$ is the index determined in Step~6. Update the value $F(\mathcal{C})$ as $i$.
\State Go over the cycle list in Step~2 and count the active ones. Set $\mathrm{active\_cycle\_cur}$ to this new number.
\State $\mathrm{count} \gets \mathrm{count} + 1$.
\State \textbf{if} $(\mathrm{active\_cycle\_cur} < \mathrm{active\_cycle\_prev})$ and $(\mathrm{count} < \mathrm{RB})$ \textbf{then}
\State \hspace{2ex} $\mathrm{active\_cycle\_prev} \gets \mathrm{active\_cycle\_cur}$.
\State \hspace{2ex} Go to Step~4.
\State \textbf{end if}
\State Construct the parity-check matrix $\mathbf{H}_{\textup{MD}}$ of the GD-MD code according to (\ref{md matrix}) and set the number of cycles-$k$ in $\mathbf{H}_{\textup{MD}}$ to $M\cdot \mathrm{active\_cycle\_cur}$.
\State \textbf{return} $\mathbf{H}_{\textup{MD}}$, the number of cycles-$k$ in $\mathbf{H}_{\textup{MD}}$, and the number $\mathrm{count}$ (i.e., the number of iterations).
\end{algorithmic}
\label{algo: FL-AO}
\end{algorithm}

Some remarks about FL-AO Algorithm are:
\begin{enumerate}
\item The algorithm takes a probability-distribution matrix as an input and performs random relocations based on this matrix (see Step~1). In particular, we feed Algorithm~\ref{algo: FL-AO} with a near-optimal distribution matrix that is provided by Algorithm~\ref{algo: MD-GRADE6}. We find a matrix (of same size as $\mathbf{P}$) with integer entries which has a concatenation vector of minimal distance to the one of the scaled matrix $\gamma \kappa \mathbf{P}$, i.e., $\gamma \kappa \mathbf{p}^{\textup{con}}$, and use it to perform the initial relocation. Due to the nature of random relocations performed initially, in every run of the FL-AO algorithm, one starts in a different point in the (globally-intractable) set of all possible relocations, and hence can end up with a different outcome in practice. All of such outcomes, however, will be better in terms of cycle counts than starting initially with no relocations, and converging to them in Algorithm~\ref{algo: FL-AO} will be faster.
\item In Step~4, in case there are multiple circulants with the same highest number of active cycles-$k$, a random selection is performed. This is due to the fact that there are no a priori criteria to distinguish them.
\item In Step~6, in case there are multiple indices regarding the relocation of the circulant of interest in Step~7, a random selection is performed. This is an essential decision in the FL-AO algorithm because the output GD-MD code might not have the minimal number of cycles (of length $=$ girth) from one algorithm run. Distinct choices lead to different paths in a tree of choices that can be traced in the algorithm, which enable lower cycle counts with multiple runs of Algorithm~\ref{algo: FL-AO}. This crucial point is addressed in \cite{homa-lev}, and we perform a random selection motivated by their observation.  
\end{enumerate}

\begin{remark} Note that the final relocation matrix produced by the FL-AO algorithm can slightly differ in relocation percentages from the output of the MD-GRADE algorithm since the final MD density is not imposed as a constraint in the FL-AO algorithm. If we have strict decoding-latency constraints, the final MD density can be inputted to the FL-AO algorithm and used as a third condition to terminate the algorithm in Step~10.
\end{remark}

\section{Experimental Results and Comparisons} \label{sec: experiment}

In this section, all GD-MD codes are constructed by FL-AO Algorithm with initial relocations based on the probability-distribution matrix obtained by MD-GRADE Algorithm. We use SC codes constructed via discrete optimization (the OO-CPO approach) \cite{oocpo} and via gradient descent (the GRADE-AO approach) \cite{GRADE}, as well as a topologically-coupled (TC) code from \cite{GRADE} as the underlying SC codes. Iterations here are always FL-AO iterations, and rates are always design rates. We first provide our results, then discuss the takeaways, and finally compare our codes with those in the literature. 

\begin{remark} Let SC Code 1 and SC Code 2 be two SC codes with coupling lengths $L_1 <  L_2$ and girth $6$, where they share the same non-zero part of a replica. Then, the total number of cycles-$6$ in SC Code 2 is equal to
$$ \sum_{k=1}^{\chi} \frac{(L_2-k+1)}{(L_1-k+1)}N_k, $$
where $N_k$ is the number of cycles-$6$ spanning $k$ consecutive replicas, $1 \leq k \leq \chi=m+1 < L_1$, in SC Code 1. In case, their girth is $8$, $\chi$ becomes $2m+1$. We use these facts in our comparisons below.
\end{remark}

\begin{table}
\caption{Lengths and Rates of Proposed GD-MD Codes and of Their 1D Counterparts}
\vspace{-0.5em}
\centering
\scalebox{1.00}
{
\begin{tabular}{|c|c|c|}
\hline
\makecell{Code name} & \makecell{Length } & \makecell{Rate} \\
\hline
GD-MD Code~1.1 & $8{,}670$ & $0.74$ \\
\hline
SC Code~1.1 & $8{,}670$ & $0.76$ \\
\hline
GD-MD Code~2.1-2.2 & $210{,}250$ & $0.81$ \\
\hline
MD-SC (NR) & $210{,}250$ & $0.81$ \\
\hline
GD-MD Code~3.1 & $17{,}480$ & $0.81$ \\
\hline
SC Code~3.1 & $17{,}480$ & $0.83$ \\
\hline
GD-MD Code~4.1 & $ 107{,}100$ & $0.81$ \\
\hline
SC Code~4.1 & $ 107{,}100$ & $0.82$ \\
\hline
GD-MD Code~4.2 & $ 154{,}700$ & $0.81$ \\
\hline
SC Code~4.2 & $ 154{,}700$ & $0.82$ \\
\hline
GD-MD Code~5 & $43{,}350$ & $0.75$ \\
\hline
TC Code~1.1 & $43{,}350$ & $0.76$ \\
\hline
\end{tabular}}
\label{table_2}
\vspace{-0.5em}
\end{table}

In the following, SC Code~x is the underlying SC code of GD-MD Code~x.y, and SC Code~x.y is the 1D counterpart of GD-MD Code~x.y, i.e, SC Code~x.y has the same replica $\mathbf{R}_1$ as SC Code~x and has the same length as GD-MD Code~x.y. Our code parameters and results are as follows.
\par SC Code~1 is an SC code with parameters $(\gamma,\kappa,z,L,m) = (4,17,17,10,1)$ and girth $6$. SC Code~1.1 is similar to SC Code~1, but with $L=30$. SC Code~1 is obtained by updating the lifting matrix of the relevant code (with the same parameters) in \cite{oocpo} constructed by the OO-CPO approach (see Section~\ref{acknowledgement}). Its partitioning and lifting matrices are given in Section \ref{sec: appendix}. SC Code~1.1 has $79{,}917$ cycles-$6$. GD-MD Code~1.1 with $M=3$ has final MD density $\mathcal{T}_1=33.82\%$ and is obtained after $12$ iterations. It has $6{,}375$ cycles-$6$. GD-MD Code~1.1 has length $8{,}670$ bits and rate $0.74$. SC Code~1.1 has length $8{,}670$ bits and rate $0.76$. 
\par
SC Code~2 is an SC code with parameters $(\gamma,\kappa,z,L,m) = (4,29,29,50,19)$ and girth $8$, constructed by the GRADE-AO approach \cite{GRADE}. For GD-MD codes in this  paragraph, $M=5$. GD-MD Code~2.1 and GD-MD Code~2.2 have final MD densities $\mathcal{T}_2=22.41\%$ after $13$ iterations and $\mathcal{T}_3=28.45\%$ after $13$ iterations, respectively. They have $3{,}180{,}285$ and $2{,}768{,}485$ cycles-$8$, respectively. GD-MD Codes~2.1-2.2 have length $210{,}250$ bits and rate $0.81$. For comparison, note that an MD-SC code with constituent SC Code~2 and $M=5$ has $16{,}809{,}705$ cycles-$8$ in the case of no relocation (NR).  
\par
SC Code~3 is an SC code with parameters $(\gamma,\kappa,z,L,m)=(3,19,23,10,2)$ and girth $8$, constructed by the OO-CPO approach \cite{oocpo}. SC Code~3.1 is similar to SC Code~3, but with $L=40$. SC Code~3.1 has $1{,}397{,}319$ cycles-$8$. GD-MD Code~3.1 with $M=4$ has final MD density $\mathcal{T}_4=31.58\%$ after $11$ iterations. It has $239{,}752$ cycles-$8$. GD-MD Code~3.1 and SC Code~3.1 have length $17{,}480$ bits and respective rates $0.81$ and $0.83$.  
\par
SC Code~4 is an SC code with parameters $(\gamma,\kappa,z,L,m) = (3,17,7,100,9)$ and girth $8$, constructed by the GRADE-AO approach \cite{GRADE}. SC Code~4.1 and SC Code~4.2 are similar to SC Code~4, but with $L=900$ and $L=1300$, respectively. SC Code~4.1 and SC Code~4.2 have $3{,}819{,}480$ and $5{,}530{,}280$ cycles-$8$, respectively. GD-MD Code~4.1 with $M=9$ has final MD density $\mathcal{T}_5=\%35.3$, and it has $112{,}959$ cycles-$8$. GD-MD Code~4.2 with $M=13$ has final MD density $\mathcal{T}_6=31.37\%$, and it has $92{,}001$ cycles-$8$. 
GD-MD Code~4.1 and SC Code~4.1 have length $107{,}100$ bits and respective rates $0.81$ and $0.82$. GD-MD Code~4.2 and SC Code~4.2 have length $154{,}700$ bits and respective rates $0.81$ and $0.82$.
\par
TC Code~1 is a TC code with parameters $(\gamma,\kappa,z,L,m) = (4,17,17,50,4)$ and girth $6$ introduced in \cite{GRADE}. TC Code~1.1 is similar to TC Code~1, but with $L=150$. TC Code~1.1 has $47{,}736$ cycles-$6$. GD-MD Code~5 with $M=3$ and constituent TC Code~1 has final MD density $\mathcal{T}_7=19.12\%$, and our FL-AO algorithm eliminated all cycles-$6$ after only $4$ iterations. GD-MD Code~5 and TC Code~1.1 have length $43{,}350$ bits and respective rates $0.75$ and $0.76$.

Below are some takeaways from these results, which are also summarized in Table~\ref{table_4}: 
\begin{enumerate} 
\item Our GD-MD codes offer significant reduction in the number of cycles-$6$ (cycles-$8$) that ranges between $92\%$ and $100\%$ ($83\%$ and $98\%$) compared with 1D SC/TC codes of the same lengths. More intriguingly, our GD-MD codes not only offer remarkably lower cycle counts compared with their 1D counterparts, but also offer lower cycle counts compared with their underlying SC/TC codes, which have notably lower lengths.

\item Our GD-MD codes outperform relevant MD-SC codes constructed in \cite{homa-lev} in terms of cycle counts. Starting with SC Code~1, our GD-MD Code~1.1 has $6{,}375$ cycles-$6$, surpassing the relevant MD-SC code designed by \cite[Algorithm~2]{homa-lev}, which has $9{,}078$ cycles-$6$ (it should be noted that the lifting matrix is somewhat different). Similarly, GD-MD Code~3.1 has $239{,}752$ cycles-$8$, whereas the relevant MD-SC code designed in \cite{homa-lev} has $249{,}320$ cycle-$8$, with the same $M$ and the same underlying SC code as our GD-MD code.

\item The comparison of cycle counts of GD-MD Code~2.2 and MD-SC NR-setting code of the same length is a demonstration of the fact that MD relocations provide us with significant reduction of cycle counts compared with the no relocation (NR) case. Moreover, the comparison of GD-MD Code~2.1 and GD-MD Code~2.2 shows that as the relocation percentage increases, the cycle count drops as expected (see Item~2 after Remark~\ref{rmk4} in Section~\ref{sec: forecast}). The design of these long codes is of interest as their finite-length performance approaches the belief propagation threshold.

\item The FL-AO algorithm narrows down the search space within the entire space of possible relocation arrangements. This is reflected in the proximity of the initial and final MD densities of GD-MD codes produced, which differ by $5-15\%$ from each other. It is also reflected in the quick convergence on a final GD-MD code by the FL-AO algorithm.

\item Despite that the final MD densities are in the range $25\%-35\%$, the number of iterations is always below $20$. This indicates that when we start with initial relocation arrangement based on a locally-optimal probability-distribution matrix, the FL-AO algorithm terminates after a moderate number of iterations, and GD-MD codes are constructed in a computationally fast manner. In fact, \cite[Algorithm~2]{homa-lev} (with no relocations initially) terminates at MD density $21\%$ in $\geq 21$ iterations for a specific MD-SC code due to the expanding/trimming steps of their algorithm. To design our GD-MD Code~3.1, which has the same parameters, our FL-AO terminates after $11$ iterations only, yielding fewer cycles-$8$. We note that our algorithm might need multiple runs since each run results in a distinct GD-MD code with a different cycle count (analogous to the tree-branch selection in \cite{homa-lev}). 

\item Since the partitioning and lifting matrices of the underlying SC codes are designed in a way that minimizes the number of cycles (of length $=$ girth), the final number of cycles we obtain after FL-AO Algorithm is lower than the upper bound on the expected number of cycles obtained in Theorem~\ref{thm: forecast}. For GD-MD Code~3.1, Theorem \ref{thm: forecast} gives a good indicator of the outcome of FL-AO Algorithm, where the (rounded) expected number of cycles-$8$ is $228{,}070$ and the actual final number of cycles-$8$ is $239{,}752$. Similarly, for GD-MD Code~2.2, the expected number of cycles-$8$ is $2{,}681{,}009$, its upper bound is $4{,}324{,}208$, and the actual final number of cycles-$8$ is $2{,}768{,}485$, falling in between, but quite close to the estimate provided by the aforementioned expected number. This observation further demonstrates the value of Theorem~\ref{thm: forecast}. However, the expected number of cycles-$8$ of GD-MD Code~4.1 is $396{,}434$ and the actual number of cycles-$8$ obtained is $112{,}959$, illustrating the strength of FL-AO Algorithm. 
\end{enumerate}

\begin{table}
\caption{Comparison of the Population of Cycles of Interest in Proposed GD-MD Codes With Their 1D-SC Counterparts and/or Those in the Literature}
\vspace{-0.5em}
\centering
\scalebox{1.00}
{
\begin{tabular}{|c|c|}
\hline
\makecell{Code name} & \makecell{Cycle-$6$ count} \\
\hline
GD-MD Code~1.1 & $6{,}375$ \\
\hline
Esfahanizadeh \textit{et al.} \cite{homa-lev} & $9{,}078$ \\
\hline
SC Code~1.1 & $79{,}917$ \\
\thickhline
GD-MD Code~5 & $0$ \\
\hline
TC Code~1.1 & $47{,}736$ \\
\hline
\hline 
\makecell{Code name} & \makecell{Cycle-$8$ count} \\
\hline
GD-MD Code~2.2 & $2{,}768{,}485$ \\
\hline
MD-SC (NR) & $16{,}809{,}705$ \\
\thickhline
GD-MD Code~3.1 & $ 239{,}752$ \\
\hline
Esfahanizadeh \textit{et al.} \cite{homa-lev} & $249{,}320$ \\
\hline
SC Code~3.1 & $1{,}397{,}319$\\
\thickhline
GD-MD Code~4.1 & $112{,}959$ \\
\hline
SC Code~4.1 & $3{,}819{,}480$  \\
\thickhline
GD-MD Code~4.2 & $92{,}001$ \\
\hline
SC Code~4.2 & $5{,}530{,}280$ \\
\hline 
\end{tabular}}
\label{table_4}
\vspace{-0.5em}
\end{table}

\section{Conclusion}\label{sec: concl}

In this work, we developed a probabilistic framework based on the GD algorithm to design MD-SC codes. We attempt to minimize the number of short cycles (of length $=$ girth) in the Tanner graph of such codes. We defined a probability-distribution matrix that characterizes the MD code design. We expressed the expected number of detrimental cycles in the GD-MD code in terms of probabilities in the distribution matrix. With this expected number in hand, we are able to answer two major questions, which will be of value to the coding theory community. The first is about an estimate of the percentage relocations required to remove all instances of a cycle in case it can be removed entirely. The second is about the best that can be done regarding a cycle given a pre-specified maximum relocation percentage, imposed by decoding latency requirements. Our MD-GRADE algorithm provides us with a locally-optimal distribution matrix that guides the FL-AO by reducing its search space. We demonstrated that, when fed with a locally-optimal probability-distribution matrix, the FL-AO can converge on excellent finite-length MD-SC designs (GD-MD codes) in a notably fast manner. Our GD-MD codes achieve notable reduction in the number of short cycles. Future work includes simulating the codes and extending the analysis to objects that are more advanced than short cycles.

\section{Appendix} \label{sec: appendix}

See Fig. \ref{fig: 2} and Fig. \ref{fig: 4} for the partitioning and lifting matrices of SC Code~1 and SC Code~3, respectively. See also Fig. \ref{fig: 3} and Fig. \ref{fig: 5} for the initial probability-distribution matrices inputted to the FL-AO algorithm and the output relocation matrices of GD-MD Code~1.1 and GD-MD Code~3.1, respectively.

\begin{figure*}
\centering
\resizebox{0.435\textwidth}{!}{\begin{tabular}{|c|c|c|c|c|c|c|c|c|c|c|c|c|c|c|c|c|}
\hline
0& 1& 1& 1& 1& 1& 1& 0& 0& 0& 0& 0& 0& 1& 1& 1& 0 \\
\hline
1& 0& 1& 1& 1& 1& 1& 0& 1& 1& 0& 0& 0& 0& 1& 1& 1 \\
\hline
1& 1& 0& 0& 1& 1& 0& 1& 1& 1& 1& 1& 1& 0& 0& 0& 1 \\
\hline
0& 0& 0& 0& 0& 0& 1& 1& 1& 1& 1& 1& 1& 1& 0& 0& 1 \\
\hline
\end{tabular}}
\vspace{5pt}

\resizebox{0.475\textwidth}{!}{\begin{tabular}{|c|c|c|c|c|c|c|c|c|c|c|c|c|c|c|c|c|}
\hline
0&  0& 12&  0& 16&  0&  7&  0&  0& 15&  0&  0&  0&  0&  9&  3& 0 \\
\hline
7&  0&  0&  5&  0&  0&  0& 15&  0&  0&  0&  0& 11&  2&  0&  0& 3 \\
\hline
0&  7& 11&  1&  0&  9&  3&  0&  0& 11& 13&  0&  0&  0&  0& 13& 0 \\
\hline
0&  0&  0&  0&  0&  0&  0& 10&  4&  0&  0& 15&  9&  5&  0&  0& 0 \\
\hline
\end{tabular}}
\caption{Partitioning matrix (top) and lifting matrix (bottom) of SC Code~1 with parameters $(\gamma,\kappa,z,L,m) = (4,17,17,10,1)$.}
  \label{fig: 2} 
\end{figure*}

\begin{figure*}
\[ \begin{bmatrix}
\text{\footnotesize 0.3672} & \text{\footnotesize 0.0664} & \text{\footnotesize 0.0664} \\
\text{\footnotesize 0.3672} & \text{\footnotesize 0.0664} & \text{\footnotesize 0.0664} 
\end{bmatrix} \]

\centering
\resizebox{0.435\textwidth}{!}{\begin{tabular}{|c|c|c|c|c|c|c|c|c|c|c|c|c|c|c|c|c|c|c|}
\hline
0& 0& 0& 0& 0& 0& 2& 1& 1& 1& 0& 2& 0& 1& 0& 0& 0 \\
\hline 
0& 0& 1& 1& 0& 0& 0& 2& 2& 0& 0& 0& 0& 0& 0& 2& 0 \\
\hline 
1& 0& 0& 1& 2& 1& 2& 0& 0& 0& 0& 0& 2& 0& 1& 0& 0 \\
\hline 
0& 0& 0& 1& 2& 0& 0& 0& 0& 0& 0& 0& 1& 2& 0& 0& 1 \\
\hline 
\end{tabular}}
\caption{Initial probability-distribution matrix (top) and relocation matrix (bottom) of GD-MD Code~1.1 with constituent SC Code~1.}
  \label{fig: 3}
\end{figure*}

\begin{figure*}
\centering
\resizebox{0.48\textwidth}{!}{\begin{tabular}{|c|c|c|c|c|c|c|c|c|c|c|c|c|c|c|c|c|c|c|}
\hline
0&1&1&0&1&2&0&2&2&0&1&1&0&1&2&0&2&2&2 \\
\hline
1&0&0&1&0&0&1&0&0&2&2&2&2&2&1&2&1&1&1 \\
\hline
2&2&2&2&2&1&2&1&1&1&0&0&1&0&0&1&0&0&0 \\
\hline 
\end{tabular}}
\vspace{5pt}

\resizebox{0.6\textwidth}{!}{\begin{tabular}{|c|c|c|c|c|c|c|c|c|c|c|c|c|c|c|c|c|c|c|}
\hline
21&0&16&3&19&1&0&0&21&5&0&0&1&0&9&0&16&1&0 \\
\hline
0&11&7&3&4&5&6&7&8&9&10&11&12&13&14&15&16&17&18 \\
\hline
0&17& 0& 6&8&10& 12& 14& 16& 18& 20& 22&  1&  3&  5& 19&9& 11& 13\\
\hline
\end{tabular}}
\caption{Partitioning matrix (top) and lifting matrix (bottom) of SC Code~3 with $(\gamma,\kappa,z,L,m)=(3,19,23,10,2)$.}
  \label{fig: 4} 
\end{figure*}

\begin{figure*}
\[ \begin{bmatrix}
\text{\footnotesize 0.24924} & \text{\footnotesize  0.02853} & \text{\footnotesize 0.02702} & \text{\footnotesize  0.02853} \\
\text{\footnotesize 0.241} & \text{\footnotesize 0.0315} & \text{\footnotesize 0.0293} & \text{\footnotesize 0.0315} \\
\text{\footnotesize 0.24924} & \text{\footnotesize 0.02853} & \text{\footnotesize 0.02702} & \text{\footnotesize 0.02853} \\
\end{bmatrix} \]

\centering
\resizebox{0.48\textwidth}{!}{\begin{tabular}{|c|c|c|c|c|c|c|c|c|c|c|c|c|c|c|c|c|c|c|}
\hline
0& 1& 0& 3& 0& 0& 0& 0& 3& 0& 0& 0& 0& 1& 3& 3& 0& 0& 0 \\
\hline
3& 0& 0& 1& 2& 0& 1& 1& 0& 0& 0& 0& 0& 0& 0& 0& 0& 0& 0 \\
\hline
0& 0& 2& 0& 0& 0& 0& 0& 0& 1& 2& 0& 2& 0& 1& 1& 0& 0& 2 \\
\hline
\end{tabular}}
\caption{Initial probability-distribution matrix (top) and relocation matrix (bottom) of GD-MD Code~3.1 with constituent SC Code~3.}
  \label{fig: 5} 
\end{figure*}

\section{Acknowledgements} \label{acknowledgement}

The authors would like to thank Christopher Cannella for providing his improved circulant power optimizer, which helped us in the design of SC Code~1 and GD-MD Code~1.1. This work was supported in part by the T\"{U}B\.ITAK 2232-B International Fellowship for Early Stage Researchers.


\end{document}